# Revisiting intrinsic spin defects in hexagonal boron nitride with r$^2$SCAN

Petros-Panagis Filippatos[1], Tom J. P. Irons [1], Katherine Inzani[1,*]

1 - *School of Chemistry, University of Nottingham, Nottingham, NG7 2RD, UK*

**Abstract**

Hexagonal boron nitride (hBN) is a wide band gap, van der Waals material that is highly promising for solid-state quantum technologies as a host of optically addressable, paramagnetic spin defects. Intrinsic and extrinsic point defects provide a range of emission energies, but the atomic-level structures related to observed transitions are not fully characterised. In this work, intrinsic point defects in bulk hBN are modelled using density functional theory at the level of the meta-generalized gradient approximation (meta-GGA), considering their formation energies, electronic spectra and magnetic properties. The meta-GGA exchange–correlation functional r$^2$SCAN is found to offer a balance between accuracy and computational efficiency for specific properties, while its predictive performance for bound-exciton stability is limited when compared to higher-level hybrid functionals. This implies opportunities for its use in optimised, hierarchical computational defect screening workflows. Under revised criteria, $V_B^-$, $B_N^0$, $B_i^+$ and $N_i^+$ defects are identified as stable colour centres with zero-phonon emission within technologically desirable wavelengths, making them promising for use in quantum networks and sensors.

Corresponding author: *  katherine.inzani1@nottingham.ac.uk

# Introduction

Quantum technologies rely on encoded quantum bits which enable applications in computing, communication and sensing. Paramagnetic colour centres in semiconductors are among the most promising candidates for solid-state qubits [1–3], where their operation as single-photon sources or spin-photon interfaces enables the generation and distribution of entangled photons in quantum networks [3–5]. Key characteristics of optically addressable spin qubits are a zero-phonon line (ZPL) in the telecom range for compatibility with optical-fibre technology, small zero-field splitting (ZFS) to enable coherent spin-state manipulation by microwave probe, and a mechanism for optical spin state initialisation and read-out via intersystem crossing (ISC), with minimal losses in the phonon sideband. To support this, the host material should have a wide band gap, and the system should have minimal spin–phonon coupling [6]. The negatively-charged nitrogen–vacancy ($NV^-$) centre in diamond largely meets these criteria, providing an optically addressable defect state that can be coherently manipulated with microwave radiation, with ISC pathways largely arising from Herzberg-Teller effects [7–9,10]. However, the $NV^-$-centre has a ZPL at approximately 2 eV and, like all candidates in three-dimensional (3D) materials, faces uncontrolled defect formation during fabrication making their deterministic placement in the host material highly challenging [11,12].

Van der Waals (vdW) materials in bulk crystals and as exfoliated two-dimensional (2D) layers offer unique advantages as qubit hosts, including reduced decoherence pathways due to a low-dimensional nuclear spin bath, minimal surface dangling bonds and alternative isotopic purification routes, contributing to spin coherence times up to tens of milliseconds [13–15]. Their topology enables precise defect placement using techniques such as scanning tunnelling microscopy or focused ion irradiation – a key requirement for scalable quantum architectures [16]. Moreover, vdW materials are ideally suited for heterostructure engineering which provides additional means for optimisation. Hexagonal boron nitride (hBN) is a leading candidate vdW host due to its wide band gap (~6 eV) [17] and chemical stability. It exhibits numerous optically active defects with bright, narrow ZPL emission even at room temperature, making it a front-runner for solid-state single-photon sources [18,19,20] and offering alternatives to the NV⁻-centre in diamond for quantum sensing [21]. Recent advances include optically detected magnetic resonance (ODMR) from individual colour centres and spin-dependent fluorescence from ensembles [22,23].

Defects in hBN exhibit emission lines ranging from 1.6 eV to 2.2 eV [20], with additional observations at higher energies of 4.1 eV and 5.3 eV. Despite extensive efforts, the microscopic origin of many emitters in hBN remains an unresolved and debated issue [24]. Room-temperature ODMR measurements have identified the negatively-charged boron vacancy ($V_B^-$) as a likely spin defect responsible for emission at 1.6 eV [25], while high-energy emitters at 4.1 eV [26] and 5.3 eV [27] have been linked to deep donor/acceptor states of $C_B$ and $C_N$ impurities. Other experimental studies have proposed different ZPLs originating from various intrinsic and extrinsic point defects, including boron vacancies ($V_B$) [19], nitrogen vacancies ($V_N$) [28], and substitutional carbon ($C_B$, $C_N$) [29] and oxygen ($O_N$) [22], while for the monolayer, defect complexes such as $V_N C_B$ [30], $C_N V_B$ [28] and $V_B$ passivated by two oxygen atoms ($V_B 2O$) [31] or three hydrogen atoms ($V_B 3H$) [32] have been reported.

Computational studies, particularly those employing hybrid exchange–correlation functionals within density functional theory (DFT) and many-body approaches, have gone some way to elucidate potential defect emitters and identify promising new candidates. Benedek *et al.* [33] proposed symmetric carbon tetramer complexes ($C_4 N$ and $C_4 B$) in bulk hBN as potential qubits with high-spin ground states and optical transitions in the visible-range. Ivady *et al.* [18] performed a detailed multireference electronic structure analysis of $V_B^{-1}$, reporting a triplet ground state with a ZPL at 1.78 eV and ZFS of 3.49 GHz, and revealing the potential for optical initialization via ISC processes. Ganyecz et al. [34] investigated the negatively-charged nitrogen interstitial ($N_i^-$) as a candidate for the experimentally observed blue emission reported in hBN, calculating a ZPL of 2.74 eV. Song et al. [35] examined the antisite defects $B_N^0$ and $N_B^+$ and proposed that these defects are responsible for the emissions at 1.58 eV and 2.63 eV, respectively. In this work the authors found $B_N^0$ to be a singlet with strong out-of-plane dipole moments while $N_B^+$ exhibited a weaker dipole and broad phonon sideband, however, the ISC mechanisms and ZFS were not extensively discussed. Overall, uncertainties remain as to which defects are responsible for particular emissions in hBN, and for many candidate defects, parameters such as the ZFS and hyperfine tensors have not been investigated. The former, which lifts the degeneracy of spin sublevels in the absence of an external magnetic field, enables coherent spin manipulation at microwave frequencies, whilst the latter can be key for characterisation by providing a fingerprint of the local electronic structure, whilst also impacting spin coherence and control.

More broadly, the generally accepted rules for ideal defect qubits – namely two defect states isolated from band edges, small spin-orbit coupling (SOC) strength and a paramagnetic ground state [6,36] – are being reconsidered in light of unconventional defects in both 3D and 2D materials having demonstrated optical initialization and readout [37,38]. These studies show that protocols involving bound-exciton defects, where an electron occupying the delocalized valence band edge is excited to an intra-gap defect state, or defects with a singlet ground state and metastable triplet state [39], can provide narrow optical transitions and spin–photon interface capabilities combined with long radiative lifetimes [40,41]. Such considerations have not yet been applied to screen intrinsic defects in hBN.

This study applies DFT at the level of the meta-generalized gradient approximation (meta-GGA) to revisit the formation energies, electronic spectra and magnetic properties of intrinsic defects in bulk hBN. Meta-GGAs include a dependence on the kinetic energy density, allowing for improved treatment of both localized and delocalized states compared to the local density approximation (LDA) and generalized gradient approximation (GGA), whilst being computationally efficient compared to hybrid exchange–correlation functionals which tend to better agree with many-body methods such as GW and experimental data [42]. The meta-GGA functionals have shown high accuracy for modelling quantum defects in 4H-SiC, with defect formation energies within ~0.2 eV of Heyd, Scuseria, and Ernzerhof hybrid functional (HSE06) results and charge transition levels (CTLs) with mean absolute errors of only ~0.18 eV relative to HSE06 [43]. Moreover, r²SCAN systematically improves the zero-phonon line (ZPL) predictions, reducing the relative error to <15% (versus >20% for PBE) [43]. Here, r²SCAN [44] is used to determine formation energies and CTLs and investigate all possible NV⁻ centre-like transitions and bound-exciton transitions arising from simple intrinsic point defects in bulk hBN. Results are systematically compared to those obtained using hybrid functionals, elucidating the role that meta-GGAs can have in optimising high-throughput, hierarchical computational defect screening workflows. In calculating the ZPL and the radiative lifetime, this study shows that r²SCAN captures key properties of the $V_B^{-1}$ triplet and further examines the potential of $N_i^{+1}$ and $B_i^{+1}$ triplet states and the $B_N^0$ singlet state for applications in quantum technologies. For stable charge states with ZPL emission in the telecom frequency band, ZFS and hyperfine tensors are evaluated and potential initialization and read-out protocols are considered.

## Methodology

**Computational details**

For DFT calculations, the Vienna Ab initio Simulation Package (VASP) [45–47] was used. All calculations were performed spin-polarized with B ($2s^22p^1$) and N ($2s^22p^3$) projector augmented wave (PAW) pseudopotentials [47,48] and the regularized-restored strongly constrained and appropriately normed (r²SCAN) meta-GGA functional [44] with DFT-D3 dispersion corrections [49]. The primitive unit cell was relaxed using a plane wave basis set with cutoff energy 750 eV and a 7 x 7 x 2 $k$-point grid, which converged total energies to within 1 meV per formula unit. Total energies were converged to $10^{-6}$ eV in the self-consistent field (SCF) loop, and structures were optimised until residual forces were less than 0.01 eV/Å. To identify the correct ground state of defect structures, the ShakeNBreak method was employed [50]. Defects were generated using the doped[51] software, in which the defect is represented within a supercell of at least 10 Å (a typical minimum size required for interactions between the defect and its periodic images to become negligible). In this work, defect calculations were performed in a non-diagonal supercell of the primitive cell, constructed via the integer transformation:

$$M = \begin{pmatrix} 2 & 2 & -5 \\ 4 & 4 & -3 \\ 1 & -1 & 0 \end{pmatrix},$$

creating a 112-atom supercell. The resulting lattice parameters were a = b = 10.88 Å and c = 10.90 Å. For the supercell calculations, a 2 x 2 x 2 $k$-point grid was used. Charge transition level diagrams were constructed with doped [51], and density of states were plotted with sumo [52].

**Defect calculations**

Formation energies and CTLs were calculated based on total energy calculations of supercells approximating the dilute limit. The formation energy of a defect, $E_{\text{form}}$, is defined as the energy difference between the defect system and its components in their reference states. For a defect $D$ with charge $q$, the formation energy is calculated as

$$E_{\text{form}}[D^q] = E_{\text{tot}}[D^q] - E_{\text{tot}}[\text{bulk}] - \sum_i n_i \mu_i + qE_F + E_{\text{corr}} \quad (1)$$

where $E_{tot}[D^q]$ is the total energy of the supercell containing the defect $D$ at charge $q$, and $E_{tot}[\text{bulk}]$ is the energy of the pristine supercell of hBN. $n_i$ is the number of atoms of type $i$ that are added ($n_i > 0$) or removed ($n_i < 0$) with chemical potentials $\mu_i$, and $E_F$ is the Fermi level relative to the valence band maximum (VBM) eigenvalue energy, which serves as a free parameter. The term $E_{corr}$ includes image charge and potential alignment corrections [53].

The chemical potentials $\mu_i$ are variables that describe the experimental synthesis conditions. In this work, $\mu_B$ is referenced to the total energy of a single atom B in a bulk crystal (α-boron), and $\mu_N$ is referenced to the total energy of a single atom N in the $N_2$ molecule in a box. $\Delta\mu_N$ and $\Delta\mu_B$ are defined with respect to these limits. To satisfy the thermodynamic equilibrium, the following condition is applied,

$$\Delta\mu_B + \Delta\mu_N = \Delta H_{form}(\text{BN}) \quad (2)$$

where $\Delta H_{form}(\text{BN})$ is the enthalpy of formation for BN, which herein is calculated as -2.84 eV in good agreement with the HSE-calculated value of -2.90 eV [54] and experimental value of -2.60 eV [55]. Equilibrium with $N_2$ represents N-rich conditions and sets an upper bound on $\mu_N$, $\Delta\mu_N = 0$, and similarly for B-rich conditions, $\Delta\mu_B = 0$; these serve as limits for experimental growth conditions.

**Transition energies, radiative lifetime and bound exciton stability**

Optical transitions in defect systems can be classified into two main categories: (i) localized-to-delocalized (LD) transitions, where one state is a localized defect level and the other is a delocalized band edge state, and (ii) localized-to-localized (LL) transitions, which occur between two defect levels situated within the band gap. To better understand the nature of these transitions, it is useful to examine how they relate to the defect energetics, CTLs and the underlying states created in the electronic structure. Following the methodology proposed by

Linderälv et al. [56], in LD transitions, absorption and emission are linked to CTLs evaluated at charge-specific relaxed geometries. The ZPL can be approximated by comparing the total energies of the two charge states in their respective relaxed atomic structures. In this case the ZPL can be approximated using the CTL position, since transitions between the localised defect levels and delocalised band edges results in the charge of the defect changing. If it is assumed that the exciton formed by transitions between localised defect levels and band edges remains localised, which is generally only realised at low temperatures, the excited-state can be modelled with constrained DFT (cDFT) [38,39] by exciting an electron from the VBM to the first unoccupied defect energy level. For the LD transition, otherwise referred to as a bound-exciton transition, the bound-exciton stability energy, defined as the difference between the ZPL and the CTL, should be positive to avoid dissociation by a charge transition process. LL transitions, on the other hand, take place entirely within band gap states. Within the cDFT approach, the total energy of the excited state configuration is calculated with one occupied energy level depopulated and a higher lying unoccupied level populated. Within this approach, the occupation of the orbitals is kept fixed during the self-consistent solution of the Kohn–Sham equations while the system is structurally optimised. This technique is widely used for excitations between two defect-localized orbitals and has shown success for the NV⁻-centre in diamond with respect to experiments [57].

To calculate the excitation lifetime for each spin- and symmetry-allowed transition, the wavefunctions were extracted after structural optimisation using a modified version of the `PyVaspwfc` code [58, 59]. The excitation lifetime is approximated using Wigner-Weisskopf theory [9,59,60]:

$$\frac{1}{\tau} = \frac{n_r (2\pi)^3 v |\bar{\mu}|^2}{3\varepsilon_0 h c^3} \quad (3)$$

where $\tau$ is the radiative lifetime, $n_r$ is the refractive index, $v$ is the transition frequency, $\bar{\boldsymbol{\mu}}$ is the k-point averaged transition dipole moment and $\varepsilon_0$ is the vacuum permittivity. $h$ and $c$ represent the Planck constant and the speed of light in a vacuum. The transition dipole moment at $k$-point $k$ is defined as:

$$\boldsymbol{\mu}_k = \frac{i\hbar}{(\epsilon_{f,k} - \epsilon_{i,k})m} \langle \psi_{f,k} | \boldsymbol{p} | \psi_{i,k} \rangle \quad (4)$$

where $\epsilon_{i,k}$ and $\epsilon_{f,k}$ represent the eigenvalues of the initial and final states, $m$ is the mass of an electron, $\psi_{i,k}$ and $\psi_{f,k}$ are the initial and final wavefunctions and $\boldsymbol{p}$ is the momentum operator.

**Zero-field splitting and hyperfine interactions**

The ZFS parameters of promising qubit candidates in their triplet ground states were calculated using first-principles methods. In general, ZFS describes the splitting between sublevels in a state with $S \geq 1$ and arises due to the combined effects of spin–orbit and spin–spin coupling. In hBN, spin–orbit coupling effects are extremely weak and the ZFS is dominated by dipolar spin–spin coupling[18]. This contribution to the ZFS was evaluated following the approach outlined in references [61–64], as implemented in the VASP code, to yield the axial ($D$) and rhombic ($E$) ZFS parameters which fully define the ZFS interaction.

The hyperfine interaction tensor describes the coupling between an electron spin $S$ and a nuclear spin $J$ and is given by:

$$A_{ij} = \frac{1}{2S} \gamma_J \gamma_e \hbar^2 \left[ \frac{8\pi}{3} \int \delta(r - R_J) \sigma(r) dr + W_{ij}(R_J) \right] \quad (5)$$

where the first term in the square brackets represents the Fermi-contact isotropic term and:

$$W_{ij}(R) = \int \left( \frac{3(r-R)_i (r-R)_j}{|r-R|^5} - \frac{\delta_{ij}}{|r-R|^3} \right) \sigma(r) dr \quad (6)$$

is the dipole–dipole anisotropic contribution. $\gamma_J$ and $\gamma_e$ represent the nuclear magneton of nucleus $J$ and the electron Bohr magneton, respectively, while $\sigma(r)$ represents the electron spin density. $R_J$ denotes the position vector of nucleus $J$ and $r$ is the electron coordinate. The subscripts *i and j* indicate Cartesian components (x,y,z) while $\delta_{ij}$ is the Kronecker delta function. After the hyperfine tensor is calculated and diagonalized, as implemented in VASP [65], the three principal values represent the hyperfine constants, which are labelled as $A_{xx}$, $A_{yy}$ and $A_{zz}$.

## Results and Discussion

### Bulk properties

Bulk hBN has the space group P6$_3$/mmc (#194) and point group D$_{6h}$ [54]. Structural relaxation with r$^2$SCAN and D3 correction results in an in-plane lattice parameter of $a$ = 2.500 Å and $c$ = 3.290 Å, in excellent agreement with experimental parameters $a$ = 2.506 Å and $c$ = 3.302 Å [66]. This slightly improves upon reported values calculated with the hybrid functionals HSE06 [54] and PBE0-TC-LRC [67], as summarized in Table 1, indicating that r$^2$SCAN with D3 correction is a good alternative for structural properties with a reduced computational cost compared to hybrid functionals [68].

The calculated band gap with r$^2$SCAN is 4.88 eV, which is underestimated by 19% compared to the experimental fundamental gap value which has been observed up to 6.00 eV [69]. This falls considerably short of the reported HSE06 value of 5.94 eV [54] and PBE0-TC-LRC value of 5.70 eV [67]. As discussed below, underestimation of the band gap by r$^2$SCAN is not crucial for the investigation of the excited states of the NV$^-$ centre-like defects. The band gap comparisons are summarized in Table 1.

Table 1. Lattice constants and electronic band gap of pristine hBN calculated using r²SCAN and, in comparison, with hybrid functionals and experiment.

|  | r²SCAN | HSE06 [54] | PBE0-TC-LRC [67] | Experiment [69] |
|---|---|---|---|---|
| Lattice constant $a$ (Å) | 2.50 | 2.49 | 2.49 | 2.506 |
| Lattice constant $c$ (Å) | 6.58 | 6.54 | 6.54 | 6.604 |
| Band gap (eV) | 4.88 | 5.94 | 5.70 | 5.50-6.00 |

**Defect energetics**

Transition level diagrams of the intrinsic interstitial, substitutional and vacancy defects in hBN calculated by r²SCAN for B-rich and N-rich conditions are shown in Figure 1. In B-rich conditions, the boron substitutional $B_N$, nitrogen vacancy $V_N$ and boron interstitial $B_i$ defects have the lowest formation energies while in N-rich conditions the nitrogen substitutional $N_B$ and nitrogen interstitial $N_i$ have the lowest formation energies.

Defect formation energies qualitatively mirror the trends reported in hybrid functional-based studies, as compared in Table 2. In particular, the energy hierarchy of neutral defect types agrees in both N-rich ($N_B < N_i < V_B < V_N < B_i < B_N$) and B-rich ($B_N < V_N < B_i < N_i < N_B < V_B$) conditions. Quantitatively, there are broad discrepancies between methods, with the difference in $E_{form}$ between r²SCAN and HSE06 [54] ranging from -0.7 % for $B_i$ to -11.6% for $V_B$ in N-rich conditions and up to -9.8% for $B_N$ in B-rich conditions, and a similar discrepancy with PBE0-TC-LRC [67] with differences from +0.7 to -6.3%. Interestingly, there is a similarly broad range of differences between the two hybrid functionals, ranging from +2% to -11%. For all available comparisons, the r²SCAN calculated values align more closely with either HSE06 or PBE0-TC-LRC values than the two hybrid functionals do with each other. While HSE06 or PBE0-TC-LRC are both range-separated functionals incorporating the same proportion of exact exchange, the latter includes a truncated Coulomb operator with a long-range correction. The two studies further differ in DFT implementation with the HSE06 study using a plane wave basis set and PBE0-TC-LRC using Gaussian-type orbitals, as well as different supercell sizes [54,67]. Overall, we find that the r$^2$SCAN formation energies for neutral defects can deviate from those calculated using HSE06 and PBE0-TC-LRC; these differences vary between defects and formation conditions, sometimes exceeding 0.5 eV. For example, the formation energy of $V_B$ in N-rich conditions predicted with r$^2$SCAN is 1 eV lower than given by HSE06, although it is only ~0.2 eV lower than the formation energy calculated with PBE0-TC-LRC under the same conditions. Similar trends were reported by Abbas *et al.* [43] in their study of NV$^-$ centre-like defects in 4H-SiC, where both SCAN and r$^2$SCAN yielded formation energies consistently lower by at least 0.5 eV relative to HSE06.

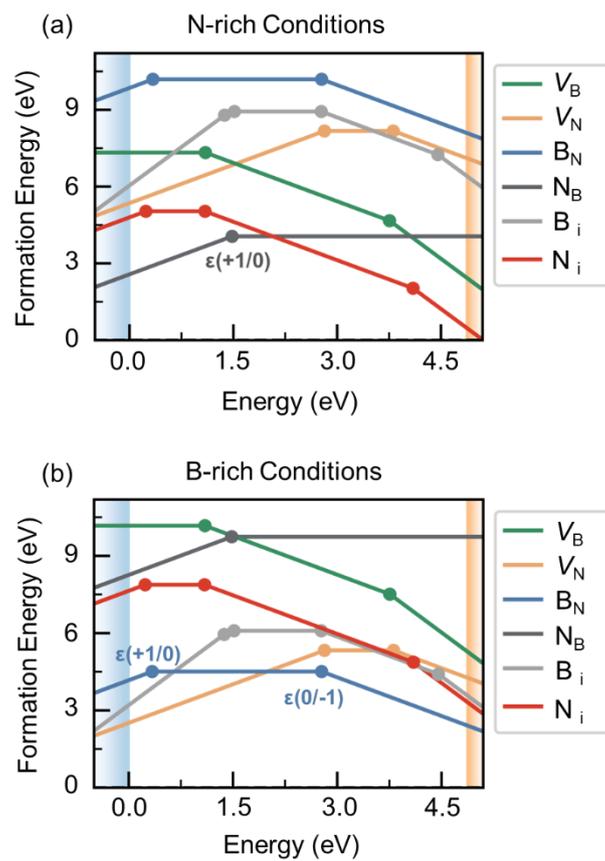

**Figure 1.** Charge transition level diagrams for intrinsic defects in bulk hBN calculated by r$^2$SCAN in the (a) N-rich limit and (b) B-rich limit.

**Table 2.** Calculated formation energies, $E_{form}$, for neutral intrinsic defects in N-rich and B-rich conditions, and percentage differences, $\Delta E_{form}$, comparing this work's r$^2$SCAN values with reported HSE06 [54] and PBE0-TC-LRC [67] values, extracted from the formation energy graphs reported in these two studies.

| Defect | $E_{form}$ (eV) | | | | | $\Delta E_{form}$(r$^2$SCAN) (%) | | |
|---|---|---|---|---|---|---|---|---|
| | N-rich | | | B-rich | | N-rich | | B-rich |
| | r$^2$SCAN | HSE06 | PBE0-TC-LRC | r$^2$SCAN | HSE06 | vs. HSE06 | vs. PBE0-TC-LRC | vs. HSE06 |
| $N_B$ | 4.057 | 4.597 | - | 9.741 | 9.836 | -11.746 | - | -0.966 |
| $N_i$ | 5.036 | 5.604 | 5.095 | 7.878 | 8.271 | -10.136 | -1.158 | -4.751 |
| $V_B$ | 7.330 | 8.334 | 7.459 | 10.172 | >10.0 | -12.047 | -1.729 | <+1.720 |
| $V_N$ | 8.173 | 8.439 | 8.650 | 5.331 | 5.911 | -3.152 | -5.515 | -9.812 |
| $B_i$ | 8.934 | 9.043 | 9.556 | 6.091 | 6.460 | -1.205 | -6.509 | -5.712 |
| $B_N$ | 10.192 | >10.0 | - | 4.508 | 5.070 | <+1.92 | - | -11.085 |

CTLs (using the notation ε(*i*/*ii*) for a transition between *i* and *ii* charge states) in relation to the band edges are shown in Figure 2, with values tabulated in Table 3 and spin multiplicities for each charge state given in Table 4. When a boron vacancy $V_B$ is formed, three $2sp^2$ and three $2p_z$ dangling bonds arise on the neighbouring nitrogen atoms (Supplementary Figure S1). $V_B$ is predicted to be stable at three charges, with CTLs ε(2-/1-) at 3.76 eV and ε(1-/0) at 1.10 eV relative to the VBM, making $V_B$ a deep acceptor. The ground state of the neutral $V_B$ has a doublet low-spin (LS) configuration, while the 1- charge is a triplet state and the 2- charge is a doublet. When a nitrogen vacancy $V_N$ is formed, three $2sp^2$ and three $2p_z$ dangling bonds arise on adjacent boron atoms (Figure S2) and give rise to gap states. $V_N$ is predicted to exist in 1+ and 1- charges, with CTLs located deep in the band gap at 3.81 eV for $\varepsilon(1-/0)$ and 2.82 eV for $\varepsilon(0/1+)$. The neutral charge has a doublet spin configuration while both 1- and 1+ charges are non-magnetic with singlet spin configurations.

Boron substitution for nitrogen, $B_N$, leads to the formation of gap states arising from the localized B–B bonds (Figure S3). Since boron has two fewer valence electrons than nitrogen, the neutral charge state results in two holes occupying the highest defect state with a non-magnetic singlet spin multiplicity. In the neutral charge, $B_N$ has occupied and unoccupied states in the band gap, making it both single donor and acceptor with CTLs at 2.76 eV for $\varepsilon(1-/0)$ and 0.34 eV for $\varepsilon(0/1+)$. The stable charges for this defect are 1+ and 1-, both of which have doublet spin multiplicities. Similarly, the nitrogen substitution for boron, $N_B$, results in states created due to N–N bonds (Figure S4). The two extra valence electrons of N fill the lowest unoccupied states, forming a singlet configuration. Removing an electron yields the 1+ charge state with a doublet spin configuration, with the charge transition $\varepsilon(0/1+)$ at 1.48 eV. All other charge states were found not to be stable.

The boron interstitial, $B_i$, has a ground state geometry intercalated between the layers of hBN (Figure S4). $B_i$ acts as a donor with the 1+ and 2+ charge states becoming stable when the Fermi level moves towards the VBM, with predicted CTLs at 1.52 eV for $\varepsilon(0/1+)$, and 1.37 eV for $\varepsilon(1+/2+)$. This defect can also act as an acceptor when the Fermi level is closer to the conduction band minimum (CBM), stabilizing the 1- and 2- charge states, with CTLs at 4.46 eV for $\varepsilon(2-/1-)$, 2.77 eV for $\varepsilon(1-/0)$. The nitrogen interstitial ($N_i$) is found to be stable at three charges, 1+, 0, 1- and 2-, with spin multiplicities of triplet, doublet and singlet respectively. In this case, $N_i$ is intercalated between the layers of hBN, sitting towards one layer in the 1-, 0 and 1+ states, with local bonding geometries highly influenced by charge state (Figure S5). For instance, $N_i^{+1}$ and $N_i^{-1}$ both form bonds with a host N atom, but the former sits atop the plane whereas the latter displaces an N atom, forming an N–N bond which straddles the plane. CTLs occur at 4.10 eV for ε(2–/1–), 1.09 eV for ε(1–/0), and 0.24 eV for ε(0/1+), reflecting that $N_i$ can function both as donor and acceptor. Weston *et al.* [54] reported that using HSE06, $N_i$ is stable as an acceptor, with the 1- and 2- charge states being stable while its positive charge states were not stable. On the contrary, Strand *et al.* [67] found that using PBE0-TC-LRC, $N_i$ can also be stabilized in the 1+ charge while the 2- charge was not stable. Interestingly, $r^2$SCAN predicts that both 1+ and 2- can be stabilized. Similar discrepancies between PBE0-TC-LRC and the other two functionals are seen in the stable CTLs for the substitutional defects (Table 3).

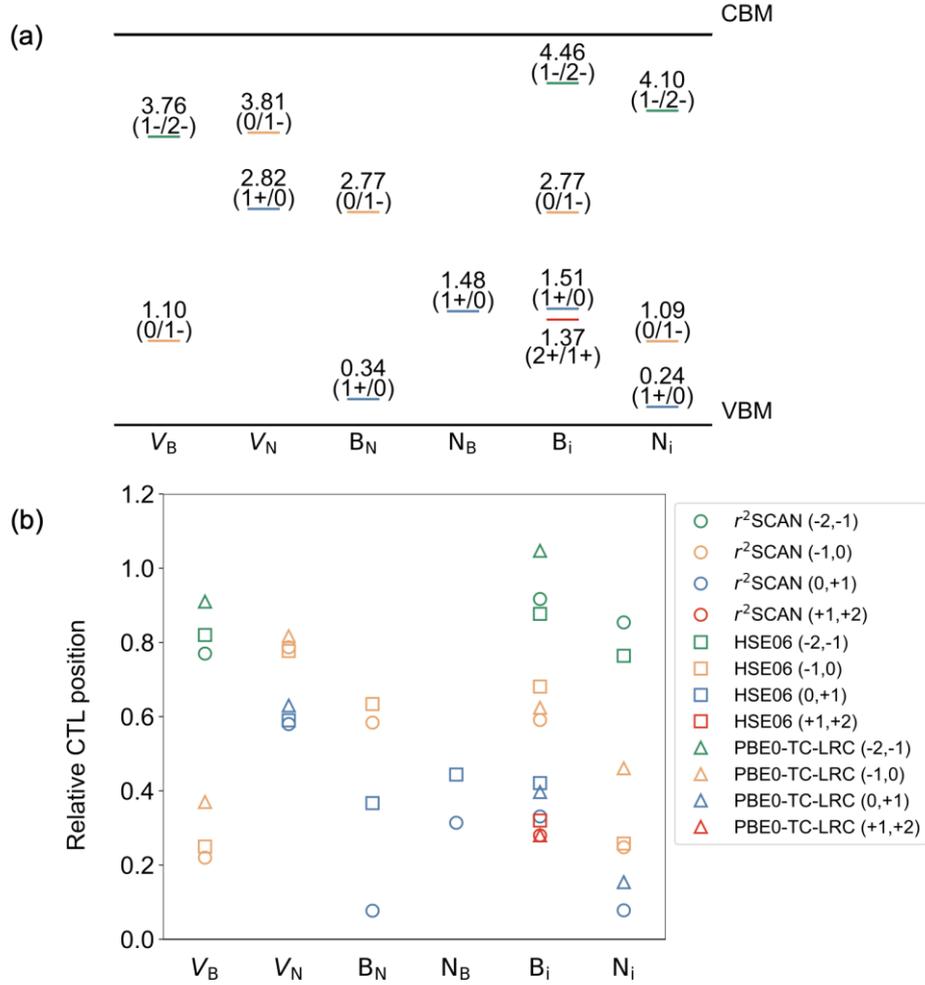

**Figure 2. (a)** Charge transition levels for intrinsic defects as calculated by r²SCAN, and **(b)** the relative CTL positions normalised by band gap for r²SCAN, HSE06 [54] and PBE0-TC-LRC [67].

These r²SCAN predictions are compared with values of CTLs reported for HSE06 [54] and PBE0-TC-LRC [67] in Table 3, and band gap-normalized CTLs, defined as $E_{rel} = \frac{E_{CTL}}{E_{gap}}$, are presented in Figure 2b. In all cases, the spin multiplets of the ground state predicted by r²SCAN agree with those given by hybrid functional approaches. Comparing normalized CTLs, r²SCAN closely reproduces most charge transition positions relative to HSE06 and PBE0-TC-LRC. For example, the ε(2-/1-) transition for $V_B$ occurs at 0.77 in the band gap compared to 0.82 for HSE06 and 0.91 for PBE0-TC-LRC. Similarly to the defect formation energies, many r²SCAN normalized CTLs (and some of the absolute CTLs) are closer to either the HSE06 or PBE0-

TC-LRC values than they are to each other. None of these approaches consistently predict higher or lower normalised CTLs with respect to the others, which is reflective of the sensitivity of diverse defect species to the methodology applied and the challenge of realising accurate calculations even within one material. In certain cases, the deviation of r$^2$SCAN CTLs from hybrid functional values would lead to a qualitatively different interpretation of the defect characteristics; for instance, r$^2$SCAN places the $\varepsilon$(0/1+) transition level for $B_N$ at 0.34 eV, suggesting a shallow donor, whereas HSE06 predicts it to be a deep donor. As defect energetics are known to be highly sensitive to choice of exchange–correlation functional, the comparable performance of r$^2$SCAN to hybrid functionals implies that meta-GGA functionals may have a role in optimising defect screening workflows, enabling the acceleration of certain steps such as the calculation of relative defect concentrations and stable charge states without compromising predictive power.

**Table 3.** Charge transition levels ($\varepsilon$(*i*/*ii*)) calculated by r$^2$SCAN and reported values using HSE06 [54] and PBE0-TC-LRC [67]. Empty cells indicate charge transitions that were not found to be thermodynamically stable for the corresponding defect under the given functional.

| Defect | $\varepsilon$(2-/1-) (eV) | | | $\varepsilon$(1-/0) (eV) | | | $\varepsilon$(0/1+) (eV) | | | $\varepsilon$(1+/2+) (eV) | | |
|---|---|---|---|---|---|---|---|---|---|---|---|---|
| | r$^2$SCAN | HSE06 | PBE0-TC-LRC | r$^2$SCAN | HSE06 | PBE0-TC-LRC | r$^2$SCAN | HSE06 | PBE0-TC-LRC | r$^2$SCAN | HSE06 | PBE0-TC-LRC |
| $V_B$ | 3.76 | 4.90 | 5.2 | 1.10 | 1.48 | 2.10 | - | - | - | - | - | - |
| $V_N$ | - | - | - | 3.81 | 4.59 | 4.60 | 2.82 | 3.48 | 3.6 | - | - | - |
| $B_N$ | - | - | - | 2.77 | 3.70 | - | 0.34 | 2.13 | - | - | - | - |
| $N_B$ | - | - | - | - | - | - | 1.48 | 2.56 | - | - | - | - |
| $B_i$ | 4.46 | 5.18 | 5.9 | 2.77 | 3.95 | 3.50 | 1.51 | 2.38 | 2.3 | 1.37 | 1.88 | 1.6 |
| $N_i$ | 4.10 | 4.47 | - | 1.09 | 1.35 | 2.50 | 0.24 | - | 0.8 | - | - | - |

Table 4. Stable defect charge states and corresponding spin multiplicities of intrinsic defects (singlet S, doublet D and triplet T).

| Defect | Charge state | | | | | Spin multiplicity | | | | |
|---|---|---|---|---|---|---|---|---|---|---|
| $V_B$ | -2 | -1 | 0 | - | - | D | T | D | - | - |
| $V_N$ | - | -1 | 0 | +1 | - | - | S | D | S | - |
| $B_N$ | - | -1 | 0 | +1 | - | - | D | S | D | - |
| $N_B$ | - | - | 0 | +1 | - | - | - | S | D | - |
| $B_i$ | -2 | -1 | 0 | +1 | +2 | D | S | D | T | D |
| $N_i$ | -2 | -1 | 0 | +1 | - | D | S | D | T | - |

**Electronic structure of selected defects**

Having determined the stable charge states for the set of simple intrinsic defects in hBN, next the most promising candidates for quantum technologies are identified by applying the criteria outlined by Weber et al [36]. Specifically, defects that exhibit a paramagnetic ground state and possess at least two distinct energy levels within the band gap to support an optical excitation are selected. In addition, these criteria are extended to consider defects with bound-exciton transitions, similar to those studied in the T-center of Si [41] and carbon-centre of $WS_2$ [38], which would be omitted by the criteria of reference 36. Of the defects identified thus far, those which satisfy these criteria are the triplet $V_B^{-1}$, singlet $B_N^0$, triplet $B_i^{+1}$ and triplet $N_i^{+1}$, for which density of states are shown in Figure 3 and Kohn-Sham energy levels in Figure 4. Symmetry analysis identifies $V_B^{-1}$, $B_N^0$, $B_i^{+1}$ and $N_i^{+1}$ as having $D_{3h}$, $C_{3v}$, $C_1$ and $C_1$ point groups, respectively. Thus, the group theory notation from their respective character tables is used for the following analysis.

For the triplet $V_B^{-1}$, the in-plane dangling bonds and out-of-plane $p_z$ orbitals of the neighbouring nitrogen atoms are responsible for the creation of non-degenerate $a$ and doubly-degenerate $e$ single particle levels in the band gap, shown in Figure 4(a). The six occupied defect levels contain 8 electrons, and the band gap is increased to 5.09 eV. The spin-down channel presents a set of unoccupied and occupied states where an optical transition can take place; using cDFT, the ZPL is calculated as 1.62 eV for the first excited state promoting an electron from the $a'_1$ level to one of the $e'_{x,y}$ degenerate energy levels, which agrees well with the experimental value of 1.65 eV [25] and values obtained using HSE06 [18]. From the low spin configuration, there are two intermediate singlet states between the ground state and the excited triplet state, with energies of 0.41 eV and 0.93 eV relative to the ground state. This finding agrees well with the computational work of Ivady et al. [18] which report the existence of two singlet states between the ground and first excited triplet states however at energies of 0.232 eV and 0.132 eV above the ground state.

For the singlet $B_N^0$, the interaction of the substitutional boron with neighbouring boron atoms forms occupied degenerate $e_{x,y}$ and unoccupied $a_1$ states located at 0.66 eV and 2.45 eV respectively, shown in Figure 4(b), in which an optical excitation can take place. The ZPL is calculated as 1.02 eV, falling into the desirable telecommunications wavelength. A triplet excited state is also identified, which is located between the ground state and the first excited singlet state. This triplet corresponds to unpairing two electrons with a single spin-flip excitation. This intermediate state is located at 0.73 eV above the ground state and corresponds to a higher energy spin configuration which can be accessed through ISC.

For the $B_i^{+1}$ defect, the band gap is reduced to 4.76 eV while occupied states are created in the spin-up channel at 1.24 eV and 1.88 eV. In the spin-down channel, unoccupied gap states are located at 3.30 eV and 4.06 eV. Following the bound-exciton protocol for this defect, for the excitation of an electron from the VBM to the first unoccupied gap state, the ZPL is 1.61 eV. Finally, for $N_i^{+1}$ the band gap is reduced to 4.50 eV, and two non-degenerate but close in energy unoccupied levels are located at 1.69 eV in the spin-down channel, with further states located close to the CBM in both spin channels. The ZPL is calculated at 0.63 eV. All the examined ZPL values are collected in Table 5.

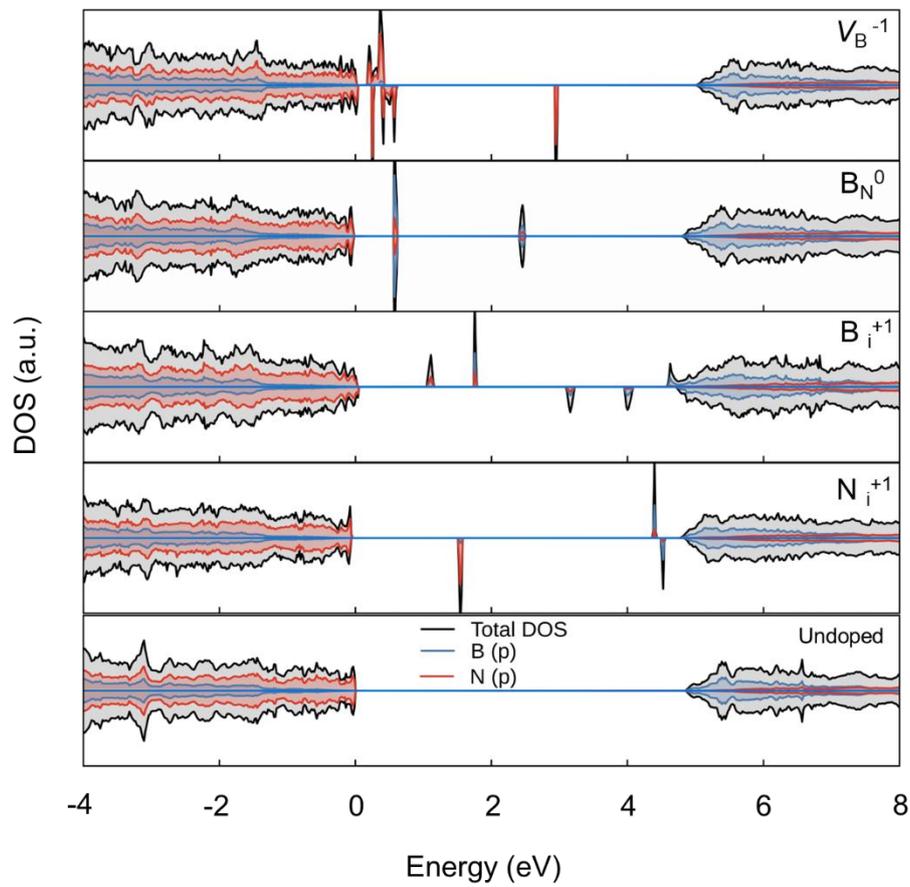

**Figure 3.** Total and orbital-resolved electronic density of states for a) triplet $V_B^{-1}$, b) singlet $B_N^0$, c) triplet $B_i^{+1}$, d) triplet $N_i^{+1}$ and e) undoped hBN. Energies are shifted such that 0 eV corresponds to the VBM.

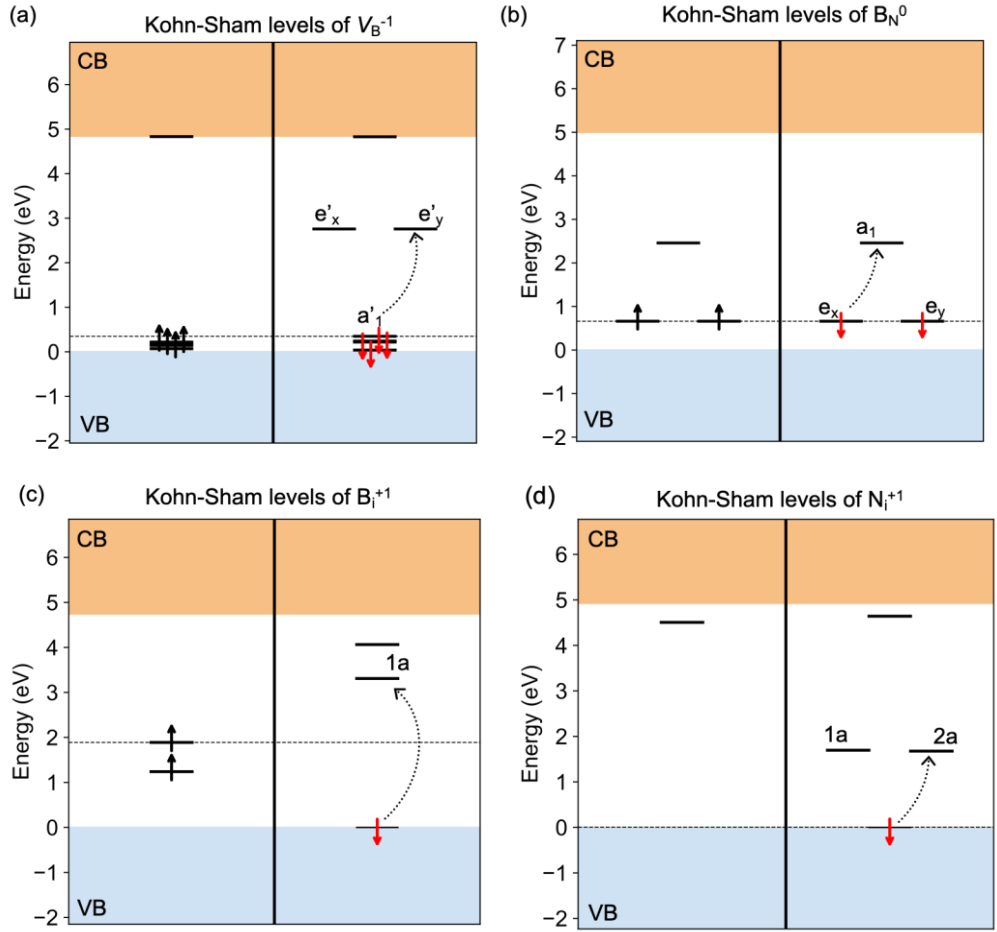

**Figure 4.** The Kohn-Sham energy levels and studied transitions for a) triplet $V_B^{-1}$, b) singlet $B_N^0$, c) triplet $B_i^{+1}$ and d) triplet $N_i^{+1}$. The black and red arrows represent the spin-up and spin-down channels respectively.

The underestimation of the band gap by $r^2$SCAN does not preclude the prediction of promising defects for use in quantum technologies, provided that the qubit is created from a localized-to-localized transition, however it can affect the description of bound-excitons which are created from delocalized-to-localized transitions. Although the relative positions of the CTLs are reasonably well-predicted with $r^2$SCAN, the absolute values are underestimated compared to those given by hybrid functionals, which indicates that the BES is not reliable. For this reason, it is crucial to calculate the BES using hybrid functionals such as HSE06 when considering bound exciton defects. Using r²SCAN we calculated the BES at 0.24 eV and 0.29 eV for $B_i^{+1}$ and $N_i^{+1}$ respectively.

**Table 5.** The band gap, ZPL and radiative lifetime computed with r²SCAN for localized-localized (LL) and localized-delocalized (LD) transitions. In this work LL transitions occur between two gap states while LD transitions occur from excitation from the VBM to the intra-gap unoccupied state.

| Defect | Transition Type | Band gap (eV) | ZPL (eV) | Excitation lifetime (ns) |
|---|---|---|---|---|
| $V_B^{-1}$ | LL | 5.09 | 1.62 | 583.50 |
| $B_N^0$ | LL | 5.12 | 1.01 | 61.88 |
| $N_i^{+1}$ | LD | 4.50 | 0.63 | 246.01 |
| $B_i^{+1}$ | LD | 4.76 | 1.61 | 79.76 |

**Qubit protocols and hyperfine splitting**

Having identified defects with their stable charges and ZPLs, the coherent manipulation of their spin states with microwave radiation along with optical initialisation and read-out of their spin states must next be considered. Coherent manipulation of the spin state by microwave radiation requires a relatively small ZFS, whilst optical initialisation and read-out requires spin-selective intersystem crossing pathways via metastable low-spin states. $V_B^{-1}$, $B_i^{+1}$ and $N_i^{+1}$ host a triplet ground state, while $B_N^0$ is a singlet; protocols for qubit operation of these is shown in Figure 5. For the triplet states, their mechanisms of operation resemble those of the well-described NV⁻-centre in diamond. The $m_s = 0$ and one of the $m_s = \pm 1$ sublevels of the triplet ground state are considered as a two-level qubit system, separated in energy by the ZFS. This is an important parameter for qubit operation as it provides useful information for the spin-lattice relaxation time [70]. The computed axial ZFS parameter (*D*) for the $V_B^{-1}$ defect is 3.1 GHz, in good agreement with the value obtained with the HSE06 functional of 3.4 GHz.[18] To our knowledge, these parameters have not yet been investigated for the other intrinsic defects in hBN. The axial ZFS parameter is calculated to be 7.5 GHz for $B_i^{+1}$ and 48.3 GHz for $N_i^{+1}$, both in their triplet ground states. The very large ZFS of $N_i$, compared to 2.87 GHz for the NV$^-$-centre [12], implies that coherent manipulation the spin state with microwave radiation would be impractical, limiting the potential of these defects as a spin qubit. We also calculated that both $B_i$ and $N_i$ have a small rhombic ZFS parameter ($E$) due to near axial symmetry, which are 0.19 GHz and 0.08 GHz respectively. Rates of non-radiative decay from the triplet excited state to a metastable singlet state are not necessarily equal for the $m_s = 0$ and $m_s = \pm 1$ components. Likewise, rates of decay from singlet metastable states to different sub-levels of the triplet ground state are not necessarily equal. This can enable optical spin polarisation of the defect, and differences in PL intensity to be observed as the relative populations of the ground state sub-levels changes. Furthermore, to manipulate the qubit, one can employ resonant microwave pulses while the readout of the qubit can be achieved by measuring the variation in fluorescence intensity.

Although most qubit protocols focus on paramagnetic ground states, it is possible to encode quantum information on a singlet ground state that has an ISC with triplet excited states. $B_N^0$ has a singlet ground state labelled as $^1A_1$, and upon illumination an electron from the occupied $e_{x,y}$ level can be excited to the empty $a_1$ level. The nearby singlet excited state $^1E$ is situated close in energy to the $^3E$ triplet state. The ZFS of this triplet excited state is calculated at 8.3 GHz, which is almost three times the value of the diamond NV$^-$-centre [9,71]. The operating principle is reversed compared to the previous cases with triplet ground states; specifically, upon excitation to the $^1E$ state, an ISC could occur to the $m_s = 0$ sublevel of the triplet state. This would enable the the qubit to be optically initialized and read out in a similar manner to that previously described for spin qubits with triplet ground states.

Calculated optical lifetimes are shown in Table 5 with $B_N^0$ having the shortest lifetime of 61.88 ns and $V_B^{-1}$ the longest lifetime of 583.50 ns. The radiative lifetime for $V_B^{-1}$ in monolayer hBN has been reported as 40 ns which is twelve times smaller than our value [72]. Similarly, the radiative lifetime for $B_N^0$ in monolayer hBN has been calculated as ~100 ns which is higher than our bulk calculations [73]. The difference between the values for the monolayer and the bulk is expected. Specifically, the radiative lifetime is highly sensitive to the dielectric environment. For example, Gao *et al* [73] reported that the $V_NN_B$ defect shows a radiative lifetime of 334 ns in a monolayer while it is reduced in the bulk to 147 ns. Thus, the apparent discrepancy simply underlines the strong dependence of the defect radiative lifetimes on the structural environment, as can also been seen in Eq. 3.

Finally, the hyperfine interactions of intrinsic defects in hBN were calculated. Understanding the hyperfine interactions is essential for mitigating decoherence but also for the precise manipulation of spin defects in hBN [74], enabling electron–nuclear entanglement for quantum memory and spin registers. The primary isotopes contributing to hyperfine interactions in hBN are $^{14}N$ and $^{11}B$ [18], possessing nuclear spins of 1 and 3/2 and gyromagnetic ratios 13.7 [75] and 3.07 [76], respectively. For $V_B^{-1}$, the hyperfine interaction with $^{14}N$ is calculated to be 34 MHz, which is underestimated compared to the experimental reported value of 47 MHz [23]. In the case of $B_N^0$, calculated hyperfine interactions with neighbouring $^{11}B$ atoms reach 140 MHz. Similarly, the $N_i^{+1}$ defect shows hyperfine interactions of 34 MHz with boron atoms, and $B_i^{+1}$ has hyperfine interactions of 28 MHz with the nearby nitrogen atoms. Hyperfine interactions play an essential role in determining the qubit coherence [77] and have been employed for quantum error correction[78] and entanglement distillation [79]. This underestimation of the hyperfine constants is likely due to the r²SCAN density functional approximation which, compared to HSE06 which incorporates partial exact exchange, will yield a more delocalised electron density due to the well-known delocalisation error arising from pure DFT exchange functionals. As can be seen in Eq. (5) and (6), the strength of hyperfine coupling is strongly dependent on the spin density around the nucleus; it thus follows that the more delocalised charge density of r$^2$SCAN will lead to a lower hyperfine coupling constant as compared to HSE06. ~~the~~

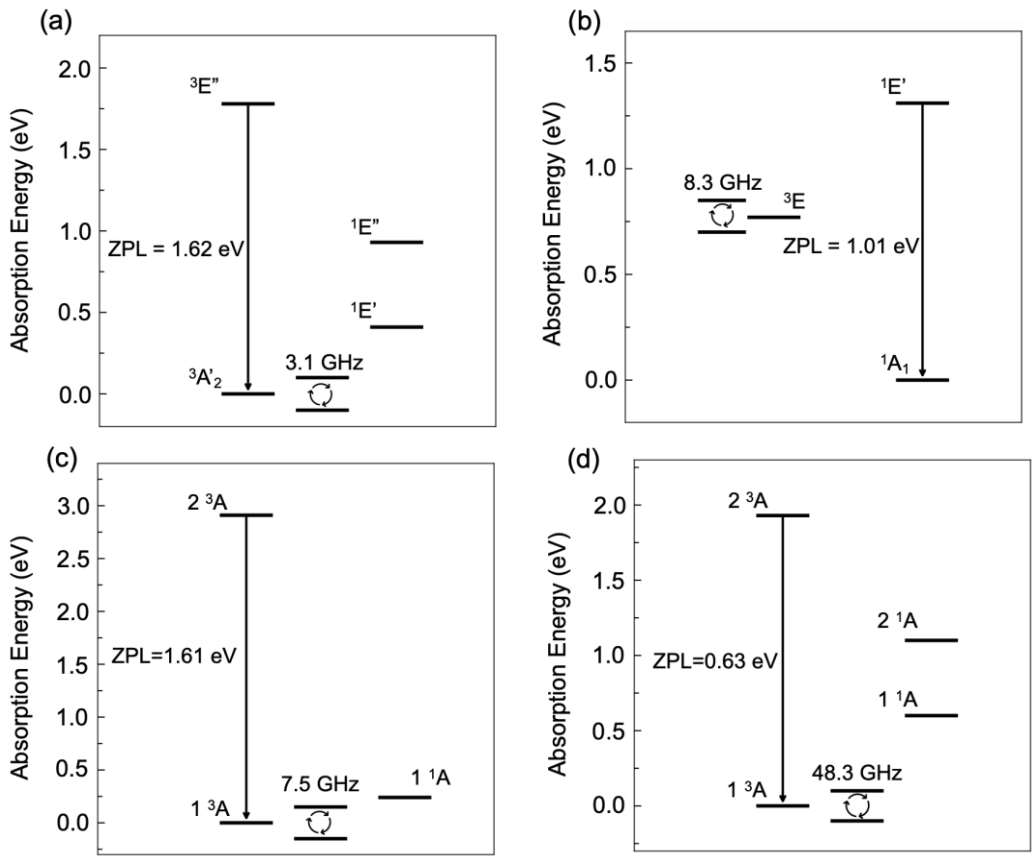

**Figure 5.** Energy level diagrams showing the ZPL and ZFS parameters, key to qubit operation for a) $V_B^{-1}$, b) $B_N^0$, c) $B_i^{+1}$ and d) $N_i^{+1}$ defect.

## Conclusions

In this work, the r$^2$SCAN meta-GGA functional was employed to systematically investigate the electronic, optical and spin properties of intrinsic defects in bulk hBN for quantum technology applications. By systematic comparison with hybrid functionals, which are considerably more computationally intensive than meta-GGAs, r$^2$SCAN was shown to provide good agreement with reported defect formation energies and charge transition levels. In capturing energetic trends of defect species and stable charge states comparable to the higher level of theory, r$^2$SCAN is suggested to be a judicious choice for high-throughput defect screening workflows. However, discrepancies in the absolute values of charge transition levels indicate that bound-exciton stabilities may be significantly underestimated.

Among all the defects studied, $V_B^{-1}$, $B_i^{+1}$ and $N_i^{+1}$ defects with triplet ground states and the $B_N^0$ with a singlet ground state were identified as the most promising for spin qubit implementation. Importantly, $V_B^{-1}$ showed excellent agreement with previously reported ZPL and ZFS values, demonstrating the validity of this methodology. $B_N^0$ showed a telecom-band ZPL and an accessible triplet state between the ground state and the first excited singlet. $N_i^{+1}$ showed potential as bound-exciton qubit with an appropriate ZPL, however its high ZFS value would preclude coherent manipulation of its spin state by microwave radiation. To advance these defect systems towards implementation, further studies are required. Specifically, the intersystem crossing rates and the spin-phonon interaction strengths should be calculated to assess spin initialization fidelity and coherence times.

## Acknowledgements

The authors are grateful to Dr Seán Kavanagh, Dr Ilya Popov, Professor Adam Gali, Dr Gergő Thiering, Dr Connor Williamson and Dr Joel Davidsson for useful discussions. T.J.P.I wishes to acknowledge support from the Royal Academy of Engineering (Grant No. CiET-2223-102). This work is supported by the UK Engineering and Physical Sciences Research Council (EPSRC) through grant No. (EP/W028131/1). DFT computations were performed using the Sulis Tier 2 HPC platform hosted by the Scientific Computing Research Technology Platform at the University of Warwick and funded by EPSRC grant No. EP/T022108/1 and the HPC

Midlands+ consortium. This work also used ARCHER2 UK National Supercomputing Services, through membership of the UK's HEC Materials Chemistry Consortium (EPSRC grant No. EP/X035859), and the University of Nottingham's HPC service Ada.

**Supplementary Information**

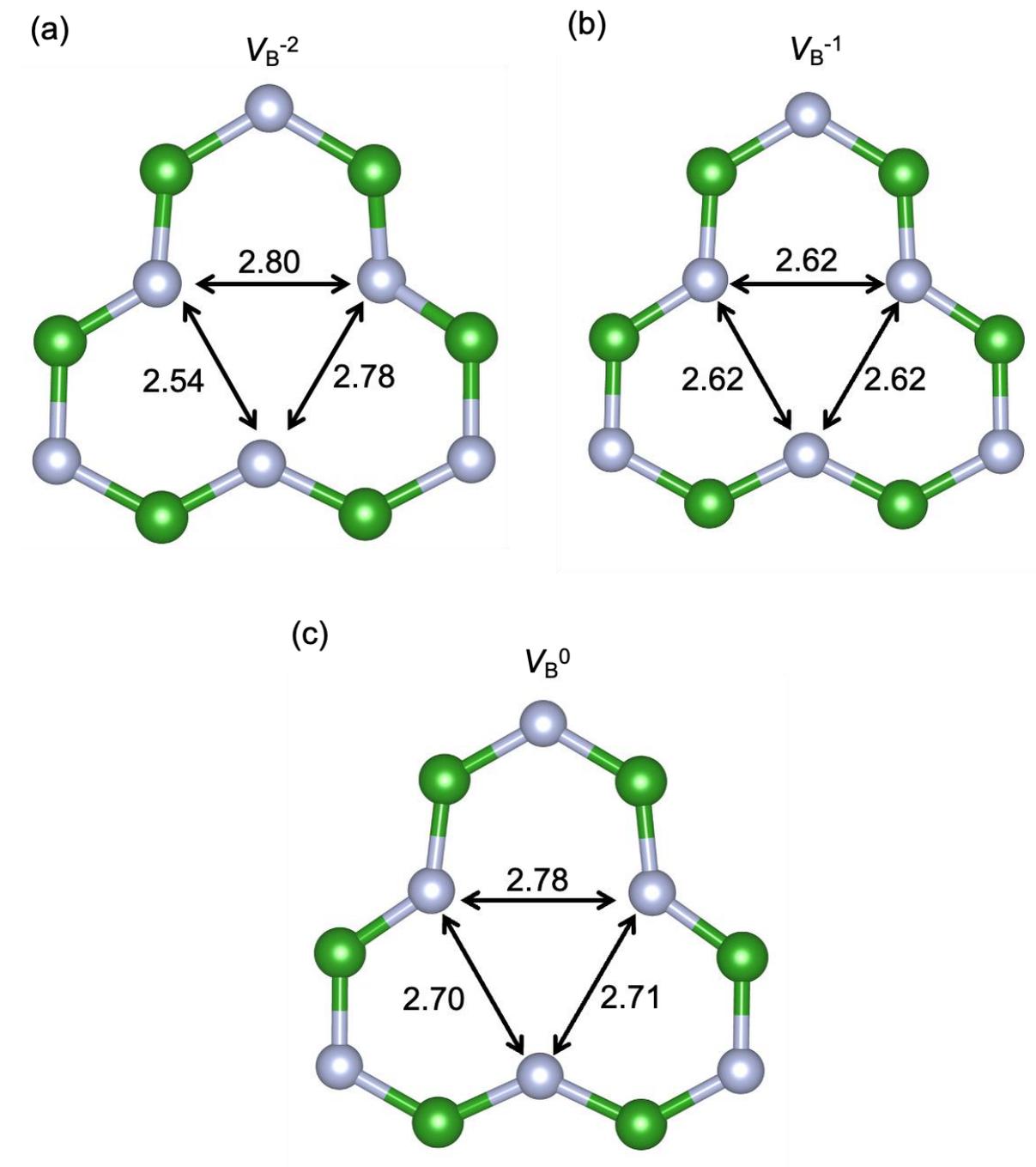

Figure S1. The B vacancy relaxed geometry in three stable charge states (a) -2, (b) -1 and (c) 0. All distances are in Å.

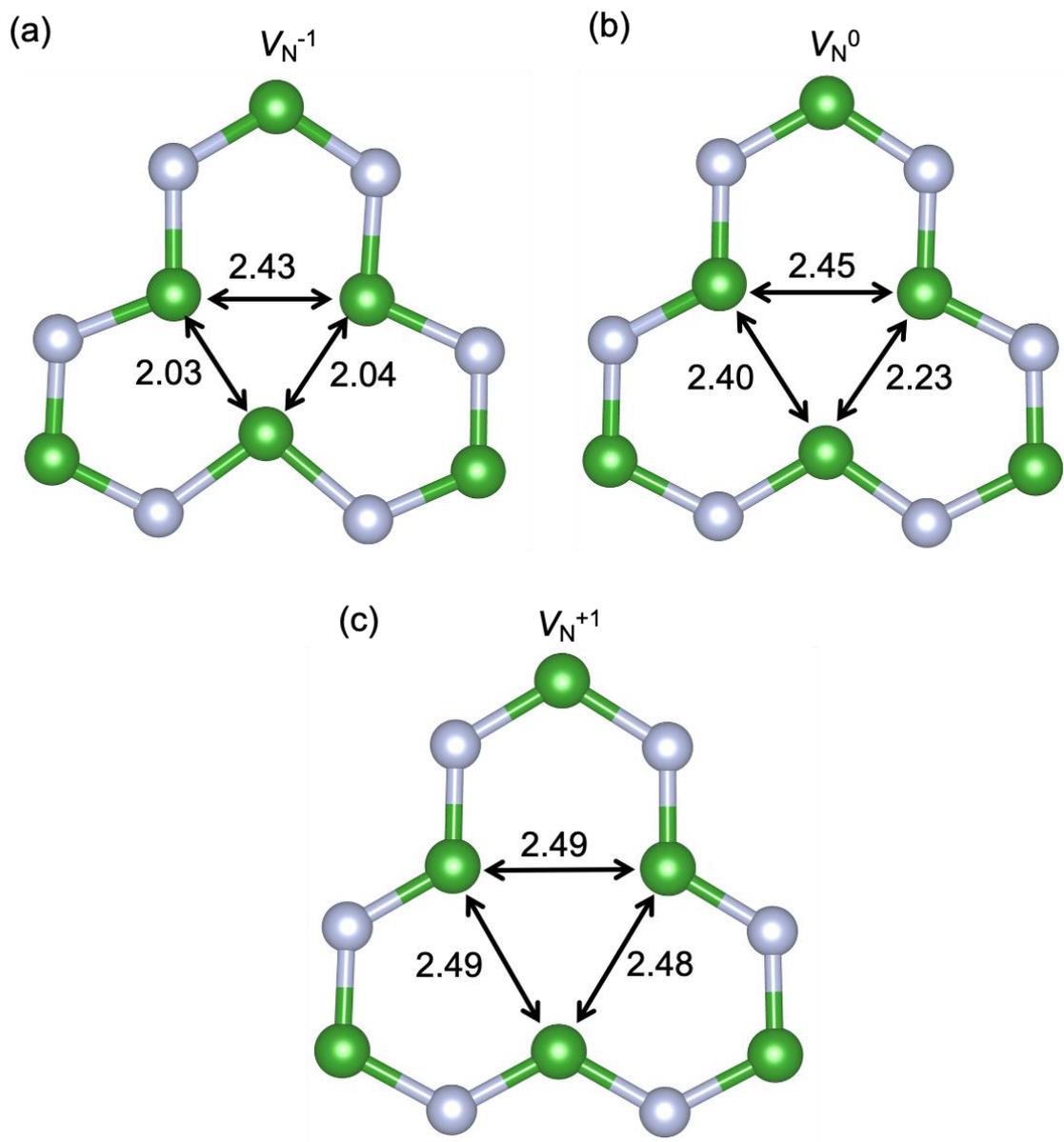

Figure S2. The N vacancy relaxed geometry in three stable charge states (a) -1, (b) 0 and (c) +1. All distances are in Å.

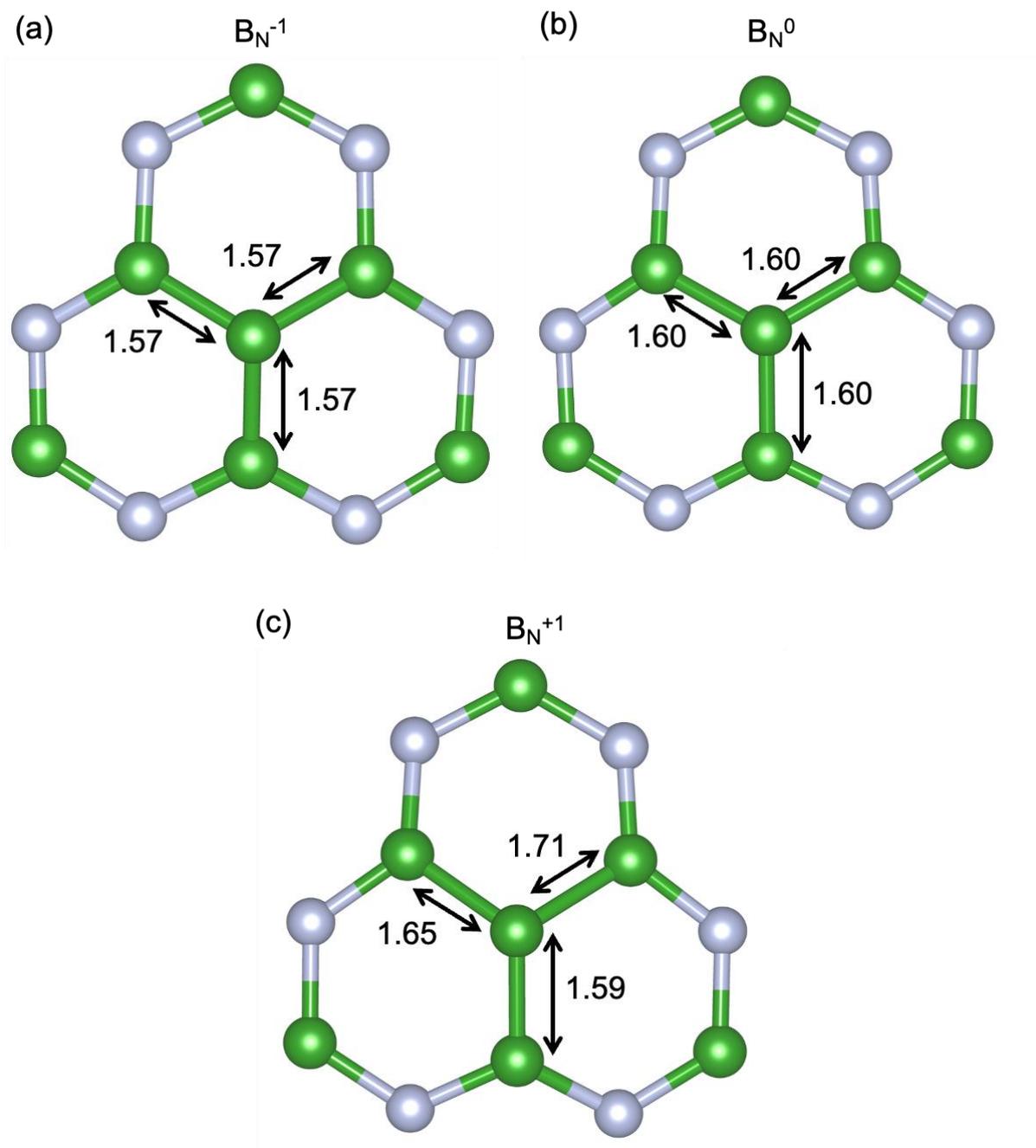

Figure S3. The B substitutional to N relaxed geometry in three stable charge states (a) -1, (b) 0 and (c) +1. All distances are in Å.

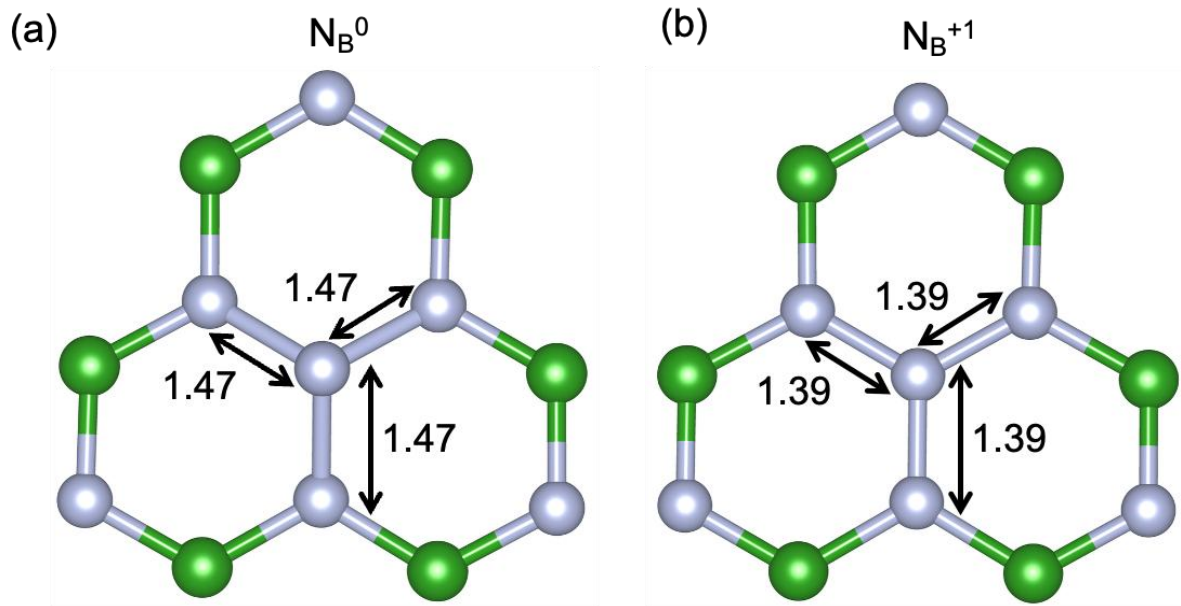

Figure S4. The N substitutional to B relaxed geometry in two stable charge states (a) 0 and (b) +1. All distances are in Å.

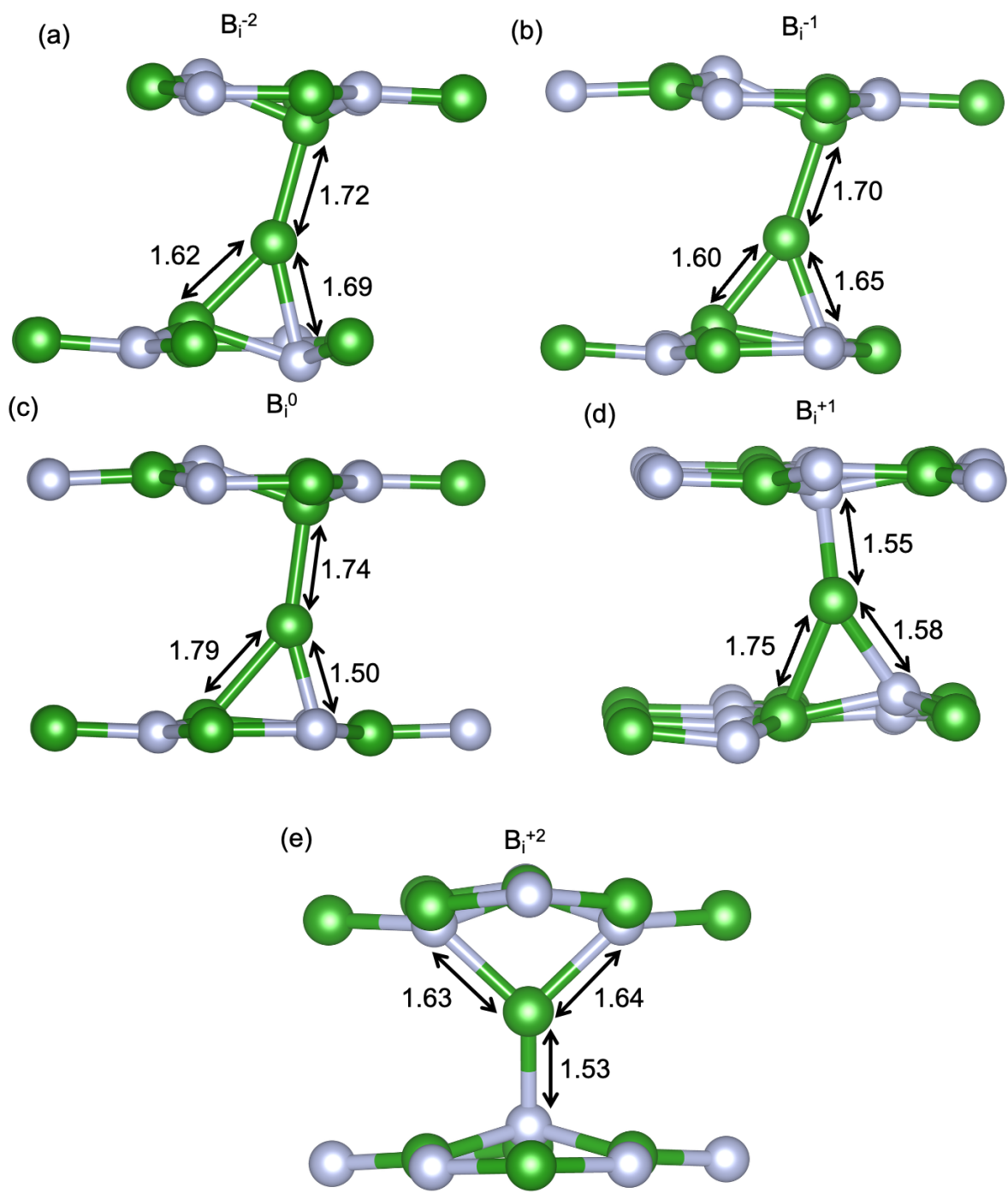

Figure S4. The B interstitial relaxed geometry in five stable charge states (a) -2, (b) -1, (c) 0, (d) +1 and (e) +2. All distances are in Å.

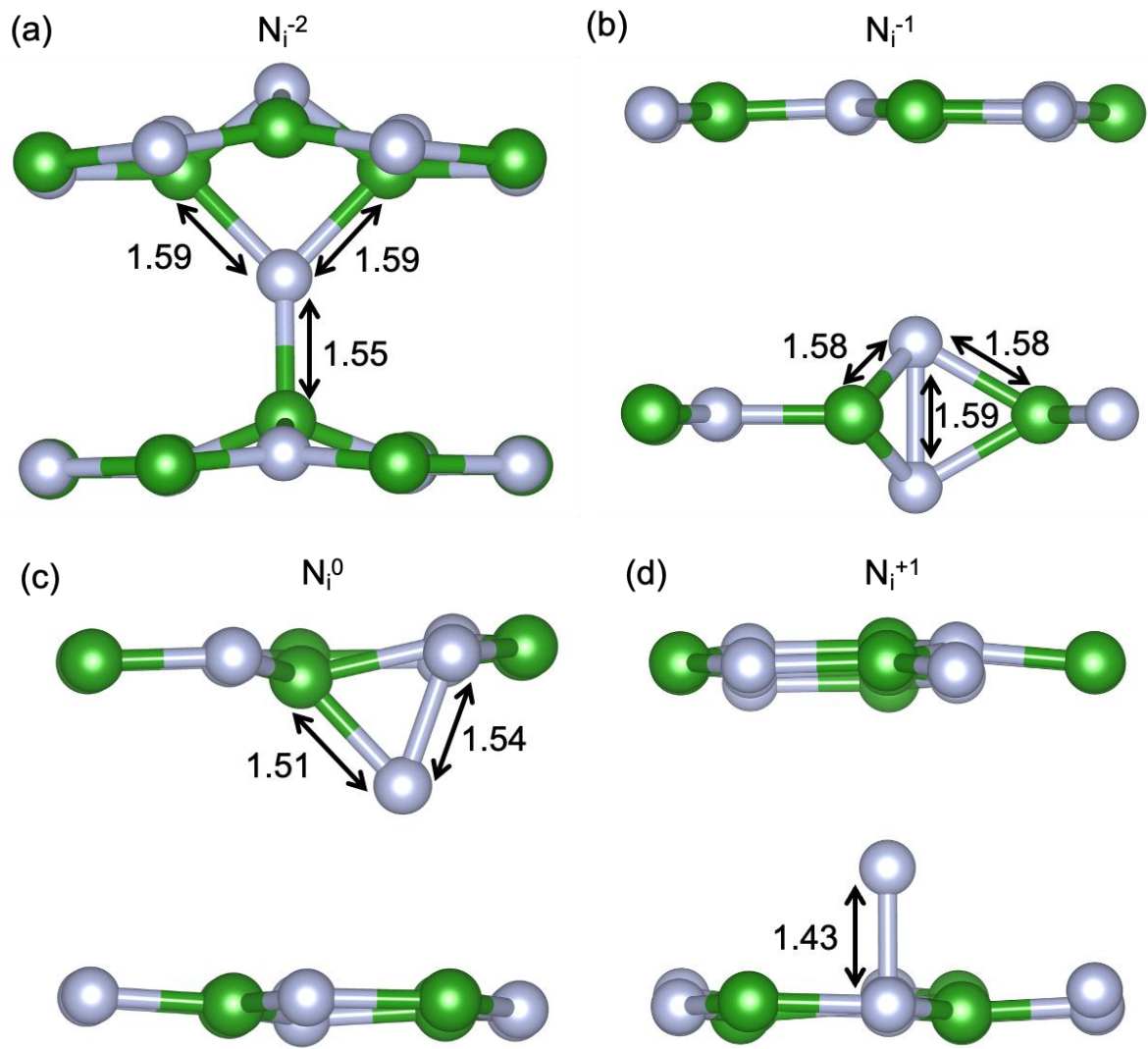

Figure S5. The N interstitial relaxed geometry in five stable charge states (a) -2, (b) -1, (c) 0 and (d) +1. All distances are in Å.